\documentclass[10pt]{article}
\usepackage{graphicx}

\begin{document}

\title{Relaxation to a Perpetually Pulsating Equilibrium}

\author{D.Lynden-Bell$^{1,2}$ \& R.M.Lynden-Bell$^{3}$ \\
\\
$^{1}$Institute of Advanced Study, Princeton, NJ 08540,USA\\
$^{2}$Institute of Astronomy, The Observatories, Madingley Road, Cambridge CB3
0HA\\
$^{3}$University Chemical Laboratory, Lensfield Road, Cambridge CB2 1EW, UK }

\date{\today}

\maketitle

\abstract{Paper in honour of Freeman Dyson on the occasion of his 80th
birthday. Normal N-body systems relax to equilibrium distributions in
which classical kinetic energy components are $1/2 kT$, but, when
inter-particle forces are an inverse cubic repulsion together with a
linear (simple harmonic) attraction, the system pulsates 
for ever. In spite of this pulsation in scale, $r(t)$, other degrees of
freedom relax to an ever-changing Maxwellian distribution. With a new
time, $\tau$ , defined so that $ r^2d/dt =d/d \tau$ it is shown
that the remaining degrees of freedom evolve with an unchanging reduced
Hamiltonian. The distribution predicted by equilibrium statistical
mechanics applied to the reduced Hamiltonian is an ever-pulsating
Maxwellian in which the temperature pulsates like $r^{-2}$. 
	
Numerical simulation with 1000 particles demonstrate a rapid relaxation to
this pulsating equilibrium. }

\section{Introduction}
	In classical mechanics when N bodies interact with forces derived from a
potential
\begin{equation} V=\sum_nV_n,
\end{equation}
where $ V_n$ is of inverse $n$th power in $|{\bf x_i-x_j}| $,
the Virial theorem reads
 
\begin{equation}
{1 \over 2}\ {d^2I \over dt^2} = 2T+\sum_n n V_n = 2E+ \sum_n(n-2)V_n.
\label{eq:2}\end{equation}
Here
\begin{equation} 
I =\sum_i m_i{\bf |x_i-\bar x|}^2=Mr^2=1/2\sum_i\sum_jM^{-1}m_im_j{\bf
|x_i-x_j|}^2,
\label{eq:3}\end{equation}
and indices $i,j$ run over the particles.
We notice that the term with $n=2$ disappears from the second sum  in equation
(\ref{eq:2}).
 Furthermore if the simple harmonic term is $ V_{-2}={1 \over 2}
\omega^2\sum_{i<j}\sum
M^{-1}m_im_j\bf|x_i-x_j|^2$ then it can be
re-expressed as $ \omega^2 I/2.$ We now specialise to the problem in
which only $V_2$ and $V_{-2}$ are present, so that any two particles of
separation $r_{ij}$ repel each other as $r_{ij}^{-3} $ and attract like
$r_{ij}$. We consider this special problem because equation (\ref{eq:2}) now
reads 
\begin{equation}
{1\over 2}{d^2I\over dt^2}=2E-2\omega^2I,
\end{equation}
so $I$ vibrates harmonically about the value $E/ 
\omega^2$. If this excitation is present initially it will continue
vibrating at the same amplitude for ever, despite the complication of the
$r_{ij}^{-3}$ repulsions of the particles.
Multiplying equation (4) by $ dI/dt $ and integrating,
\begin{equation}
{1\over 4} ({dI\over dt })^2=2EI-\omega^2I^2-M^2{\cal L}^2,
\end{equation}
where the last term is the integration constant. Using $I=Mr^2$ this
 becomes  \begin{equation}
1/2 (\dot r^2+{\cal L}^2r^{-2}+\omega^2r^2)M=E,
\end{equation}
which may be compared with the energy of a particle of mass
$M$ orbiting with specific angular momentum ${\cal L}$ under a central
simple harmonic force. Evidently
\begin{equation}
d^2r/dt^2={\cal L}^2r^{-3}-\omega^2 r.
\end{equation}
To save writing unimportant details hereafter we take all the masses
equal, so that $m_i=m=M/N$.
The equation of motion for  ${\bf x_i}$ now reads
\begin{equation}
md^2{\bf x_i}/dt^2=-m\omega^2({\bf x_i-\bar{x}})
-\partial V_2/\partial \bf{x_i},
\end{equation}
where the simple harmonic forces were combined using equation (3).
We now employ rescaled variables defined by
\begin{equation}
{\bf X_i}=({\bf x_i-\bar {x} })/r.
\end{equation}
Then
\begin{equation}
{d^2{\bf x_i} \over dt^2}={d^2r\over dt^2}{\bf X_i}+2{dr\over dt} {d{\bf
X_i}\over dt}+r{d^2{\bf X_i}\over dt^2}=({\cal L}^2r^{-3}-\omega^2r){\bf
X_i}+r^{-3}r^2{d \over dt}\left(r^2{d{\bf X_i}\over dt}\right).
\end{equation}
 Introducing a new \lq time' $\tau$ by $d/d\tau=r^2d/dt$, we
notice that ${\cal L}\tau$ is the azimuth of the particle in the imaginary
orbit introduced under equation (6). The equations of motion become
\begin{equation}
md^2{\bf X_i}/d\tau^2=-m{\cal L}^2{\bf X_i}-\partial{\cal
V}_2/\partial{\bf X_i},
\end{equation}
where ${\cal V}_2=r^2V_2$. Since $V_2$ is homogeneous of degree $-2$ in the
${\bf x_i}$, one merely replaces the ${\bf x_i}$ by ${\bf X_i}$ to make
${\cal V}_2$ from $V_2$. The result does not depend on $r$ explicitly.
Equation (11) is thus an autonomous equation for the evolution of the
reduced variables ${\bf X_i}$, but as functions of $\tau$ rather than $t$.

From their definition (9) the ${\bf X_i} $ are constrained so that both
\begin{equation}
\sum{\bf X_i} =0 \hbox{  and  }\sum{\bf X_i^2} = N.
\end{equation}
Now the $E$ in equation (1) is the energy relative to the centre of mass
since the $I$ is measured in that frame, see (3). 
Our energy equation is
therefore
\begin{equation}
E={1\over2}\sum m[d({\bf x_i-\bar x)}/dt]^2+V_{-2}+V_2={1\over2}M(\dot
r^2+\omega^2r^2)+r^{-2}[{1\over2}\sum m(d{\bf X_i}/d\tau)^2+{\cal V}_2].
\end{equation}
Eliminating $\dot r$ via (6) and multiplying by $r^2$ we obtain
\begin{equation}
{1\over2}M{\cal L}^2={1\over2}\sum m(d{\bf X_i}/d\tau)^2+{\cal V}_2.
\end{equation} 
So the "energy" of the reduced variables in $\tau$-time is 
$M{\cal L}^2/2$. Had we directly integrated equation (11) to find this
energy, it would not have been obvious that the integration constant was 
zero.

We are now in a position to state our problem in statistical mechanics.
Given that the ${\bf X_i}$ must satisfy the constraints (12), what is
their statistical equilibrium and how does any such equilibrium translate
back into the eternally pulsating variables ${\bf x_i} $ and ${\bf \dot
x_i}$? We find the equilibrium in section 2. In section 3 we demonstrate
by numerical experiment with 1000 particles that a system started well
away from that pulsating equilibrium relaxes to the predicted
ever-pulsating equilibrium. Finally in section 4 we remark on the
solutions of the corresponding problem in quantum mechanics.  

The problem is exceptional in that the normal dissipation of the basic
breathing mode via Violent Relaxation \cite{L-B} is exactly
absent. Nevertheless the other modes of the system do dissipate so Violent
Relaxation is not totally absent. Although the long range simple harmonic
force seems very artificial the net effect is exactly that found by Newton
for the gravitational force within a homogeneous body. Such bodies have
inspired many delightful studies by the great mathematicians including one 
 by Freeman Dyson \cite{Dyson}. 
	 The special case of our problem in which $V_2$ is zero is the N-body problem
exactly solved by Newton in the Principia \cite{newton}. The pulsating
equilibrium idea arose in our earlier generalisations of his work to other
extraordinary N-body problems that are exactly soluble in both classical and
quantum mechanics \cite {L-B^2a,L-B^2b}. The special case when the harmonic
force is absent and the particles are on a line is the exactly soluble
Calogero model \cite{calogero}. 

\section {Statistical Mechanics of the Reduced System}
	The constraint $\sum {\bf X_i}^2=N$ tells us that the $3N$
coordinates lie on a hypersphere of radius $\sqrt N$ 
in $3N$ dimensions. Each of the three constraints $\sum{\bf X_i}=0$ gives
a hyperplane through the hypersphere's centre so the first of them reduces
the
$3N$-sphere to a $(3N-1)$-sphere; imposing the others as well reduces
that to a ($3N-3$)-sphere. If ${\cal V}_2$ were zero the representative
point would move freely on such a hypersphere and with ${\cal V}_2$
present the motion is still a Hamiltonian one in the angles. The reduced
system (11) still has a conserved total angular momentum. Writing ${\bf
X_i'}$ for $d{\bf X_i}/d\tau$ the conservation law is 
\begin{equation}\sum {\bf
X_i}\times m {\bf X_i'}=\sum ({\bf x_i}/r)\times mr^2{d \over dt}({\bf
x_i}/r)=\sum {\bf x_i}\times m{\bf \dot x_i}={\bf J}. 
\end{equation}
Incorporating this constraint too, the reduced system has $3N-7$ degrees
of freedom. For details of the angles on the hypersphere and their
corresponding momenta see the appendix to \cite {L-B^2b}. Using Lagrange
multipliers $m\beta^*/2,{\bf \gamma},\delta$ to impose the
constraints of constant 'energy' $M{\cal L}^2/2,\   {\bf J}, $ and $  \sum
{\bf X_i}^2$ ,the equilibrium distribution function factorises into a
distribution of momenta $m{\bf X'}$ given by,
\begin{equation}
f \propto \exp[-\beta^*m{\bf X'^2}/2-{\bf \gamma.(X}\times m{\bf
X'})],
\end{equation}and one dependent on the spatial coordinates.

 Writing ${\bf \gamma} =-\beta^*{\bf \Omega}$ and omitting
a further function of position 
\begin{equation} f\propto \exp[-\beta^*m({\bf
X'-\Omega\times X})^2/2].
\label{eq:disu}\end {equation}
Returning to our original variables ${\bf x}=r{\bf X}$ and writing ${\bf
\dot x=v}$,  we have ${\bf X'}=r[{\bf v}-(\dot r/r){\bf x}]$ where $r(t)$
is the ever pulsating scale. In terms of our old variables $f$ takes the
form\begin{equation}
f \propto \exp[-\beta^*mr^2({\bf v-u})^2/2],
\label{eq:disv}\end{equation}
where ${\bf u}=(\dot r/r){\bf x -\Omega}r^{-2}\times {\bf
x}$. We see that ${\bf v}$ is distributed Maxwellianly relative to the
mean {\bf u(x},t).This mean moves with a time-dependent 'Hubble' flow
superposed on a time-dependent rotation ${\bf \Omega}r^{-2}$.
Furthermore the temperature of the distribution is time-dependent
with $\beta^*r^2$ taking the place  of the normal $\beta$ so that the
temperature is proportional to $r^{-2}$. We intentionally omitted the
spatial distribution from the above as it contains all the complications
of the problem. With both attractive and repulsive forces present we
expect phase transitions of solids to liquids and gases even in classical
physics without quantum phenomena. At low temperatures we expect a solid
lattice but it can not be perfectly regular as the spacing must increase
outward as the pressure decreases. None of this prevents the whole body
undergoing large amplitude rescaling pulsations with the associated
time-dependent rotation predicted above. At high enough temperatures 
${\cal V}_2$ is always small compared with the kinetic energy and it can
be neglected except as the means by which the system relaxes to its
pulsating equilibrium whose density is then given by \begin{equation}
\rho=\rho_{ 0} \exp-[\delta Z^2+q(X^2+Y^2)],
\end{equation}where ${\bf X}=(X,Y,Z)$. The coeficient $q$ is best expressed
in terms of a reduced omega $\omega^2=m\beta^*\Omega^2$ and takes the form
$4q=3-\omega^2+\sqrt{(3-\omega^2)^2+2\omega^2}.$ The Lagrange multiplier
$\delta$ is $\delta=q+\omega^2/2.$ Notice that $\delta \rightarrow
q\rightarrow3/2$ as
$\Omega\rightarrow0$. From equation (19) we may calculate the moment of
inertia and thence the total angular momentum is
${\bf J=\Omega}M/q$. The other Lagrange multiplier is determined from        
$ m\beta^*{\cal L}^2=3+\omega^2/q$.

\section{Simulating the Approach to the Pulsating Equilibrium}

\begin{figure}
\includegraphics[width=12cm]{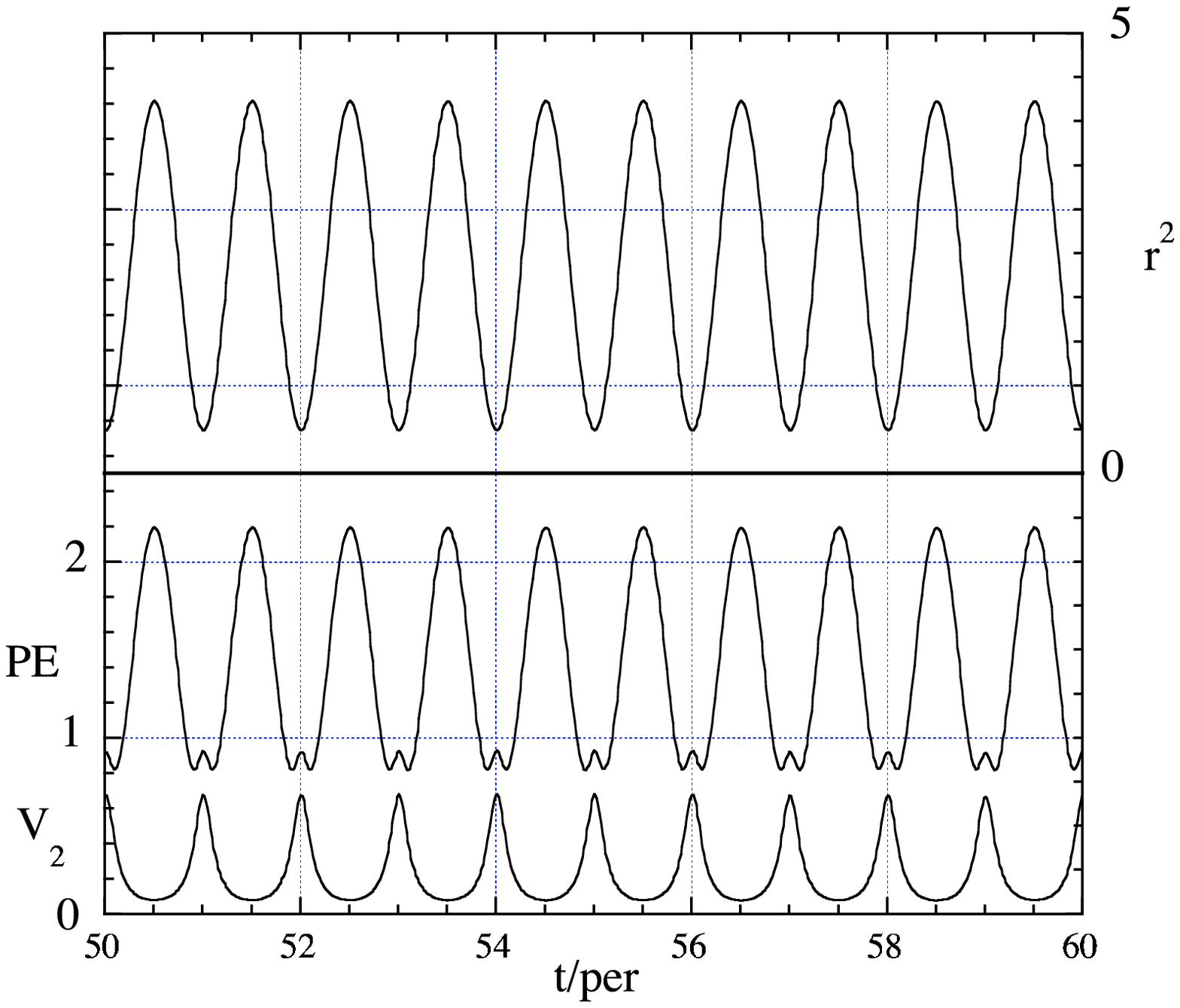}
\caption{\label{fig:en} Variation of $r^2$ (top), total potential energy
(middle) and $V_2$ (bottom) during 10 pulsations following the first 50
pulsations.}
\end{figure}

Numerical simulations were carried out on a system of 1000 particles of equal
mass with pair interactions depending on their separation $r_{ij}$,
\begin{equation}
V(r_{ij})={1 \over 2 m^2}[r_{ij}^2+r_{ij}^{-2}],
\end{equation}
using the method of molecular dynamics \cite{aandt}. The unit of  mass was
defined so that the total mass $M=1000m$ was equal to one, and the equations
of motion were integrated using the Verlet velocity algorithm with a time
step of 0.001. With this choice of parameters the period of the oscillation of
$r^2$ is $\pi$ time units. The starting configuration was constructed from a
cubic array of particles with an initial interparticle spacing of 0.1 and the
origin of the coordinates was defined as the cube centre. Velocities were
chosen randomly from a Gaussian distribution, and adjusted so that the
velocity of the centre of mass was equal to zero and the total energy $E$ was
equal to some specified value. The coordinate system was rotated so that the
angular momentum was along the $z$ direction. In order to study the approach
to equilibrium, the initial velocity and spatial distributions were perturbed
by scaling the $z$ velocities and $z$ positions by a factor of two. Such a
scaling leaves the angular momentum along the $z$-axis and unchanged,

The perpetual pulsating is  illustrated in figure \ref{fig:en}. The top curve
in this figure shows $r^2$ as a function of time measured in periods of
$\pi$ time units for periods 50-60 since the beginning of the simulation. 
The harmonic pulsation of $r^2$  is clear and both amplitude and phase are the
same as at the beginning of the simulation. The lower two curves show the
total potential energy and the contribution from $V_2$, the inverse square
term. It can be seen that the latter term is only important during the phase
of the pulsation when the system is compressed so that $r^2$ is small. 

\begin{figure}
\includegraphics[width=12cm]{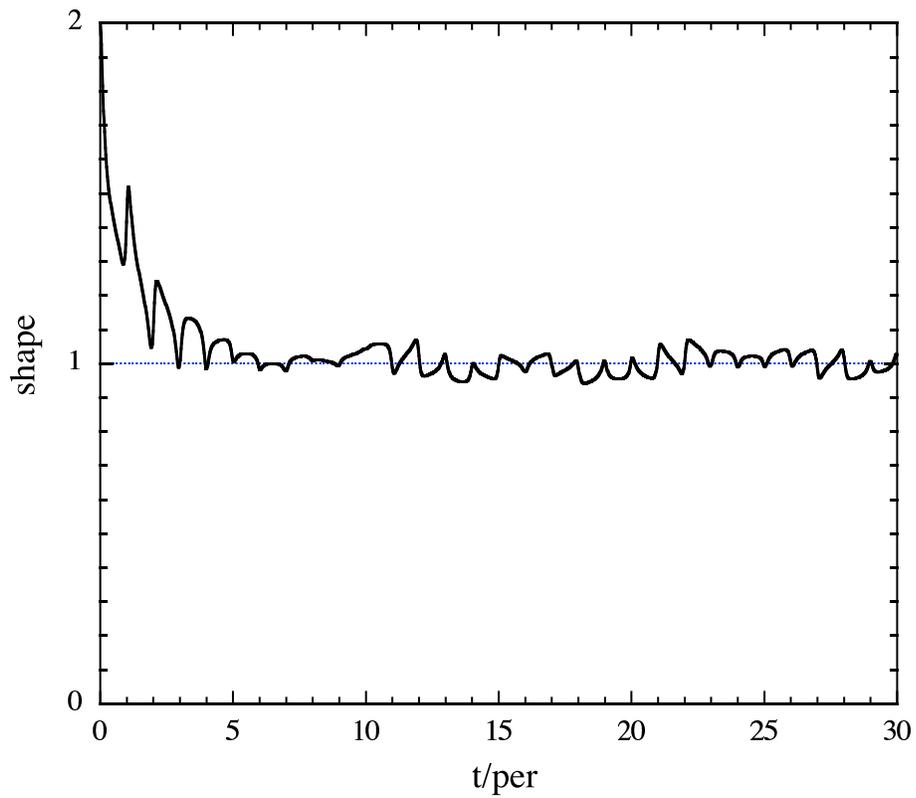}

\caption{\label{fig:shape}
Variation of the cluster shape $(\sum z^2_i/\sum x^2_i)^{1/2}$ at the
beginning of the run. Note that the initial anisotropy relaxes in about 5
pulsations}
\end{figure}

\begin{figure}
\includegraphics[width=12cm]{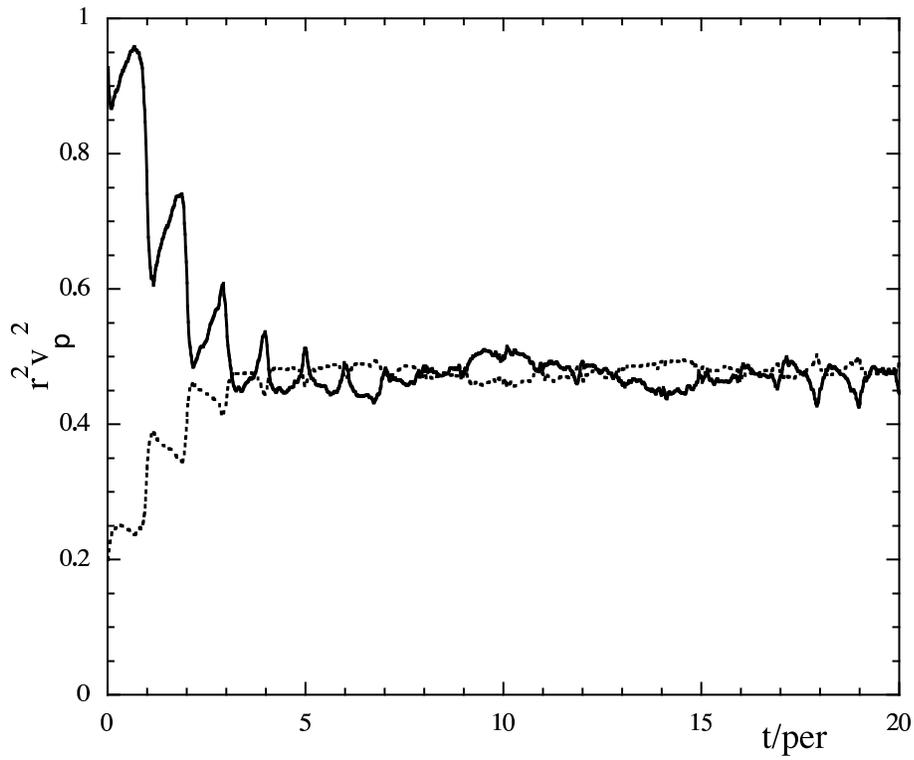}

\caption{\label{fig:r2u2} Variation near the beginning of the run of the  mean
square peculiar velocities in the $z$(solid) and $x$(dashed) directions scaled
with
$r$. Note that these relax to equal values in about 5 pulsations}

\end{figure}

The relaxation toward equilibrium of both the shape of the cluster and the
distribution of the peculiar velocities
$ ({\bf v-u)=v_p}$  towards equilibrium are shown in figures \ref{fig:shape}
and \ref{fig:r2u2}. Figure (\ref{fig:shape}) shows the shape of the cluster
as measured by the ratio
$(\sum z_i^2/\sum x^2_i)^{1/2}$, which tends to one at equilibrium, since the
only angular momentum in the simulation is the small one statistically
generated in the initial conditions. Although the system is initially far from
equilibrium, the shape relaxes over about 5 periods to an approximately
spherical distribution with equal second moments in $x$ and
$z$, although both quantities are pulsating. Fluctuations in the ratio
remain for many periods.  Figure (\ref{fig:r2u2}) shows the changes in
$r^2\sum v_{pz}^2$ and $r^2\sum v_{px}^2$ as  functions of time.
According to equation (18) these quantities should be constant and equal at
equilibrium. Relaxation of the difference occurs over about 5 pulsation
periods.

\begin{figure}
\includegraphics[width=8cm]{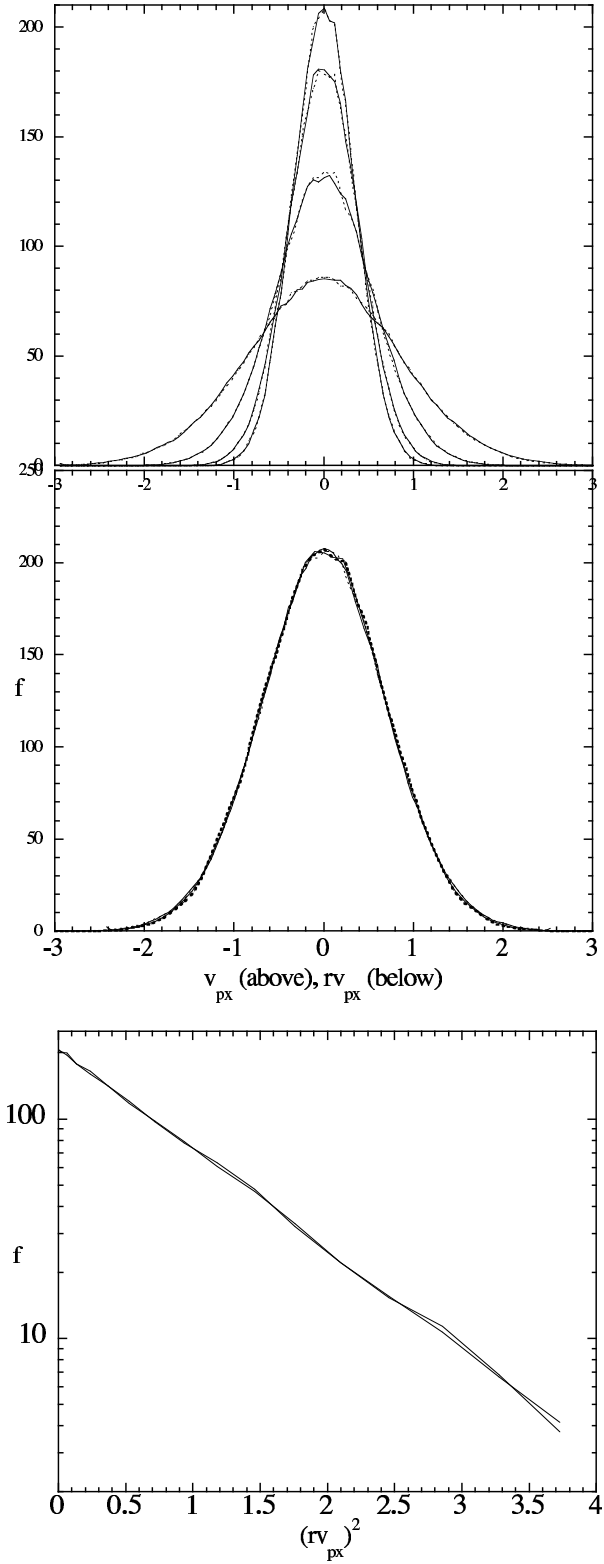}

\caption{\label{fig:gauss} The equilibrium distribution of peculiar
velocities. The upper graph shows the distribution of the peculiar velocities
for eight phases of the pulsation cycle. The broad distributions correspond to
phases when the cluster is compressed and the narrowest portions to expanded
phases. In the middle graph the velocities have been rescaled with $r$
demonstrating that the eight distributions coincide. The bottom portion
demonstrates the Maxwellian nature of the distribution by showing the
logarithm of the distribution function is linear in $r^2v_{px}^2$.}
\end{figure}

The Maxwellian distribution of peculiar velocities $({\bf v_p)}$ predicted in
equation (18) is illustrated in figure (\ref{fig:gauss}). The top part of the
figure shows the velocity distributions at eight different phases
of the pulsation. At the phase when the cluster is
compact the velocity distribution is broader than when the cluster is
extended, but in every case the distribution is Maxwellian. The middle part
of the figure verifies that the the width of the distribution is inversely
proportional to
$r$ as predicted by equation (18), as it shows that, when the velocities are
rescaled by $r$, the distributions coincide. To decrease numerical
fluctuations all these graphs average  together the distributions of the x,y
and z components of velocity relative to the predicted mean and also average
the distributions taken at the same phases of fifteen pulsations at the end
of the run. The logarithmic plot of the distribution function against $(
v_{px})^2$ is shown to be linear in the final graph, confirming the Maxwellian
form of the distribution.

\section{Conclusions}The predicted pulsating equilibrium is approached and
forms a limit cycle.
		Limit cycles are well-known in dynamics but it is more unusual to come
across a pulsating Maxwellian distribution that continues with no change of
entropy. However a related case is found in the Planck distribution
of photons which remains rescaled but unchanged in shape or entropy as the
universe expands. That is not true of the distribution function of massive
particles which need collisions to maintain equilibrium as they cease to be
relativistic.
	
In \cite{L-B^2b} we discussed the quantum mechanics of a closely related
problem including the relevent Fermi-Dirac and Einstein-Bose distributions.
The Hamiltonian separates in the form $H=\bar H({\bf \bar x})+\tilde
H(r)+r^{-2}\hat H({\bf X})$. The use of momenta allowed us to get the
solutions without the introduction of $\tau$-time but the pulsations of
the equilibrium have to be sought out in the correlations whereas they
stand out more clearly in the classical case discussed above.
\bigskip

\section{Acknowledgments}
	We are grateful to the Institute of Advanced Study for providing the
environment at Princeton where this work was done. D.L-B.  thanks the
Monell Foundation for supporting his visit. R.M.L-B. thanks the Chemical
Engineering department of Princeton University.

\suppressfloats
\pagebreak

\end{document}